\documentclass[aps,
prx,
reprint,
showpacs,
superscriptaddress]
{revtex4-2}

\usepackage[dvips]{graphicx}
\usepackage{dcolumn}% Align table columns on decimal point
\usepackage{bm}% bold mat/cite
\usepackage{xspace}
\usepackage{multirow}
\usepackage{natbib}
\usepackage{color}
\usepackage[version=3]{mhchem}

\usepackage[normalem]{ulem}

\begin{document}

\preprint{}
\title{Charge Density Waves and the Effects of Uniaxial Strain on the Electronic Structure of 2H-NbSe$_2$}
\author{Asish K. Kundu}
\affiliation{Condensed Matter Physics and Materials Science Department, Brookhaven National Laboratory, Upton, New York 11973, USA\\}
\affiliation{National Synchrotron Light Source II, Brookhaven National Laboratory, Upton, New York 11973, USA}
\author{Anil Rajapitamahuni}
\author{Elio Vescovo}
\affiliation{National Synchrotron Light Source II, Brookhaven National Laboratory, Upton, New York 11973, USA}
\author{Ilya I. Klimovskikh}
\affiliation{Donostia International Physics Center, 20018 Donostia-San Sebastian, Spain\\}
\author{Helmuth Berger}
\affiliation{Institute of Condensed Matter Physics, Ecole Polytechnique Fédérale de Lausanne, 1015 Lausanne, Switzerland}
\author{Tonica Valla}
\email{tonica.valla@dipc.org}
\affiliation{Condensed Matter Physics and Materials Science Department, Brookhaven National Laboratory, Upton, New York 11973, USA\\}
\affiliation{Donostia International Physics Center, 20018 Donostia-San Sebastian, Spain\\}

\date{\today}

\begin{abstract}
Interplay of superconductivity and density wave orders has been at the forefront of research of correlated electronic phases for a long time. 2H-NbSe$_2$ is considered to be a prototype system for studying this interplay, where the balance between the two orders was proven to be sensitive to band filling and pressure. However, the origin of charge density wave in this material is still unresolved.  Here, by using angle-resolved photoemission spectroscopy, we revisit the charge density wave order and study the effects of uniaxial strain on the electronic structure of 2H-NbSe$_2$. Our results indicate previously undetected signatures of charge density waves on the Fermi surface. The application of small amount of uniaxial strain induces substantial changes in the electronic structure and lowers its symmetry. This, and the altered lattice should affect both the charge density wave phase and superconductivity and should be observable in the macroscopic properties.
\end{abstract}
\vspace{1.0cm}

%\pacs {74.25.Kc, 71.18.+y, 74.10.+v, 74.72.Hs}

\maketitle\section*{introduction}
Ever since the discovery of charge density wave (CDW) in NbSe$_2$ and related transition metal di-chalcogenides, these materials became a model system for studying the interplay of superconductivity (SC) and phases with broken lattice translational symmetry, that can result in exotic phenomena such as Higgs mode, topological superconductivity and Cooper pair density wave \cite{Moncton1975,Mattheiss1973,Morris1975,Chu1977,Littlewood1982a,Wilson2001,Luo2016a}. However, the origin of CDW in 2H-NbSe$_2$ is still debated, with the strong electron-phonon coupling (EPC) and the nesting of segments of the Fermi surface (FS) or shallow van Hove singularities (vHS) being the main driving candidates \cite{McMillan1977,Kohn1959,Varma1983,Rice1975,Wilson2001,Borisenko2009,Flicker2016,Valla2004,Johannes2006,Zheng2018,Zhu2017,Flicker2015a,Valla2000b}. This is in part due to the fact that the spectroscopic signatures of CDW formation have been very hard to detect \cite{Hess1990,Kiss2007,Borisenko2008,Shen2008,Arguello2015}.

NbSe$_2$ has been a test system for various theoretical models describing the formation of CDW in two-dimensional (2D) materials, pointing to the shortcomings of the initial FS nesting scenario \cite{Kohn1959}. The strong momentum-dependent EPC has been later shown to be crucial for the formation of CDW while the importance of the FS nesting was diminishing with the coupling strength \cite{McMillan1977,Kohn1959,Varma1983,Valla2004}. The inelastic X-ray scattering studies on 2H-NbSe$_2$, found that the acoustic phonons softened to zero frequency over a broad momentum region around the CDW ordering vector $q_{\mathrm{CDW}}$, inconsistent with a FS nesting scenario, where sharp dips were expected \cite{Weber2011,Kohn1959}. 

Although the SC - CDW interplay has been recently shown to be sensitive to band filling, that could be altered by intercalation or gating \cite{Fan2019,Zhang2022,Xi2016}, it was realized early on that the balance between these orders is particularly susceptible to the strain or pressure \cite{Chu1977}. The basis for tunability is in the fact that the electronic hopping and vibrational properties are highly sensitive to small changes in the interatomic distances occurring under pressure. Under the increasing hydrostatic pressure, SC in 2H-NbSe$_2$ has been found to strengthen while the CDW weakens, as expected for competing orders \cite{Chu1977}.

The uniaxial strain can have much larger effect on the electronic structure and the shape of the FS than the equal amount of isotropic (hydrostatic) pressure due to the larger distortion of the unit cell as it relaxes orthogonally to the strain axis according to Poisson’s ratio. In systems with large FSs, where the Fermi wavevector $k_F$ approaches the Brillouin zone (BZ) boundary, a small uniaxial strain can transform a 2D closed FS into an open, quasi 1D one. That offers a large non-linear response and tunability of the system through a relatively small uniaxial deformations. The best example is Sr$_2$RuO$_4$, a layered material with two pairs of equivalent vHSs near the Fermi level in the relaxed state. A small uniaxial strain (0.6\%) enhances the superconducting transition temperature ($T_c$) in Sr$_2$RuO$_4$ by 230\% \cite{Steppke2017}. The \textit{in-situ} angle-resolved photoemission spectroscopy (ARPES) studies have shown that this enhancement coincides with shifting of one set of vHSs to the Fermi level under strain \cite{Sunko2019a}.  

%###################################################
\begin{figure*}[htpb]
\begin{center}
\includegraphics[width=15cm]{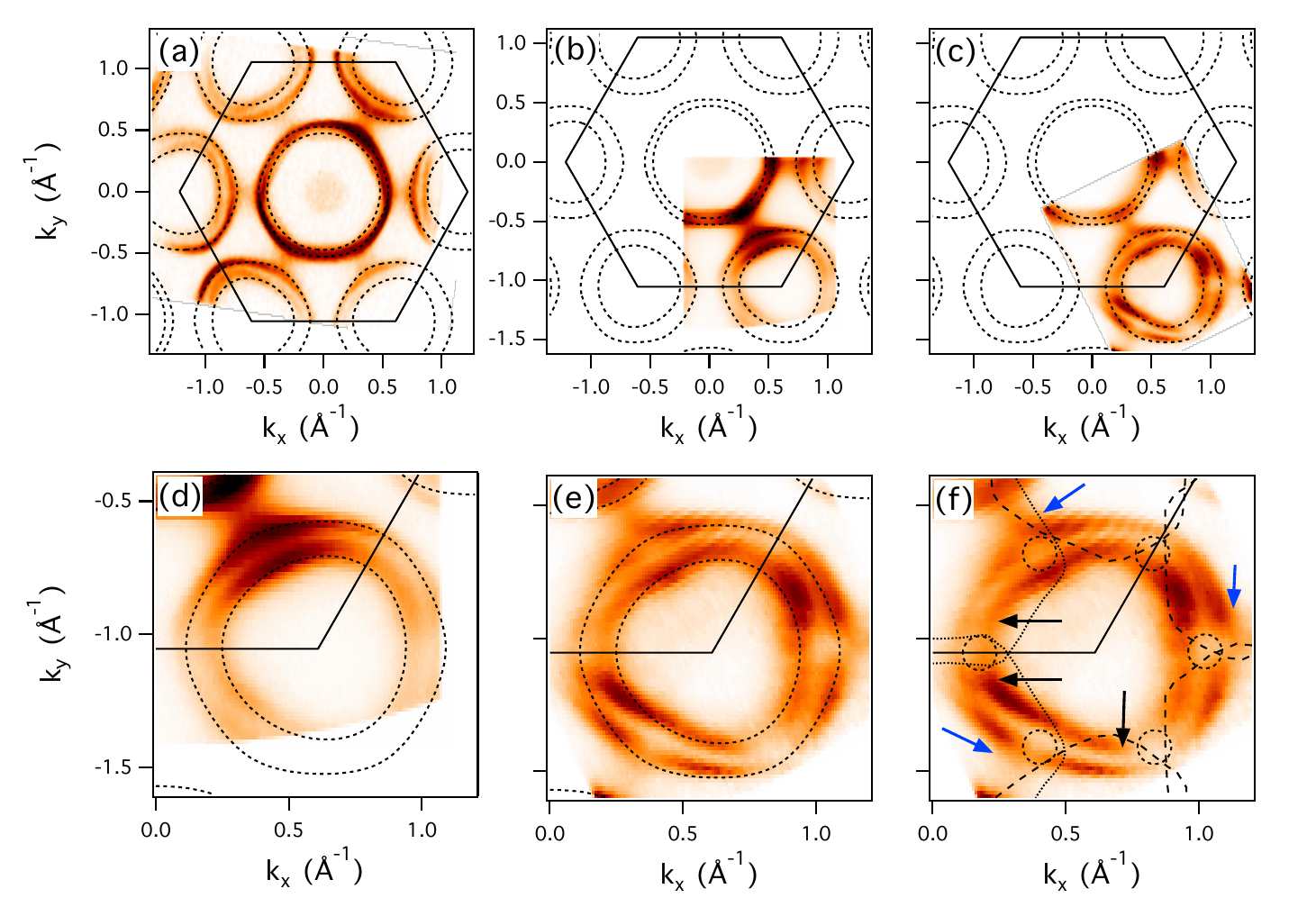}
\caption{CDW in 2H-NbSe$_2$. (a) Normal state Fermi surface (FS) of 2H-NbSe$_2$ recorded at $h\nu = 80$ eV and at $T=40$ K. Tight binding approximation for the FS is plotted by dashed lines. (b) The same FS, recorded at He I ($h\nu=21.22$ eV) and $T=35$ K. (c) The same as (b), but recorded in the CDW state at $T=18$ K. (d) Close-up of K-centered Fermi barrels from (b) (normal state). (e) Close-up of  K-centered Fermi barrels from (c) (CDW state). (f) The same as in (e), with the dashed circles indicating reduction in intensity at the Fermi level, the black arrows indicating straightening of the inner barrel and the blue arrows indicating bending of the outer barrel. Also shown are sketches of the experimental FSs displaced by $\pm q_{\mathrm{CDW}}$ in the $k_y$ direction. The outer $\Gamma$-centered barrels and the inner K-centered barrels are represented by dotted and dashed lines, respectivelly.
}
\label{Fig1}
\end{center}
\end{figure*}
%###################################################

In materials with co-existing phases, the unaxial strain can shift the balance between the phases or turn them on and off. In 2H-NbSe$_2$, recent scanning tunneling microscopy (STM) studies have shown that a small amount of anisotropic biaxial strain alters the triangular (3Q) CDW phase into a unidirectional (1Q) phase \cite{Gao2018,Soumyanarayanan2013,Wieteska2019}. Theoretical calculations  show a significant difference between the FSs for these two CDW phases and associate the observed 3Q to 1Q transition with the small amount of uniaxial strain \cite{Flicker2015,Flicker2016,Wieteska2019}. However, aside from random effects due to the local strain variations, observed in STM, the experimental studies under uniaxial strain applied in a controlled way have never been performed on 2H-NbSe$_2$ \cite{Gao2018,Soumyanarayanan2013,Wieteska2019}.

Here, we perform the high-resolution ARPES studies of the electronic structure in the relaxed and uniaxially strained 2H-NbSe$_2$. In the relaxed state, we clearly see the effects of CDW on the K-centered Fermi barrels that are consistent with the star-like displacements of the Nb atomic sites in the CDW state. In the uniaxially strained samples, we detect shifts of the vHSs along the $\Gamma-$K lines in the BZ that effectively lowers the symmetry of the electronic structure from 6-fold symmetric to a 2-fold one. This should have significant consequences on the macroscopic CDW and SC phases and should also affect the normal state properties.

%###################################################
\begin{figure}[htpb]
\begin{center}
\includegraphics[width=8.6 cm]{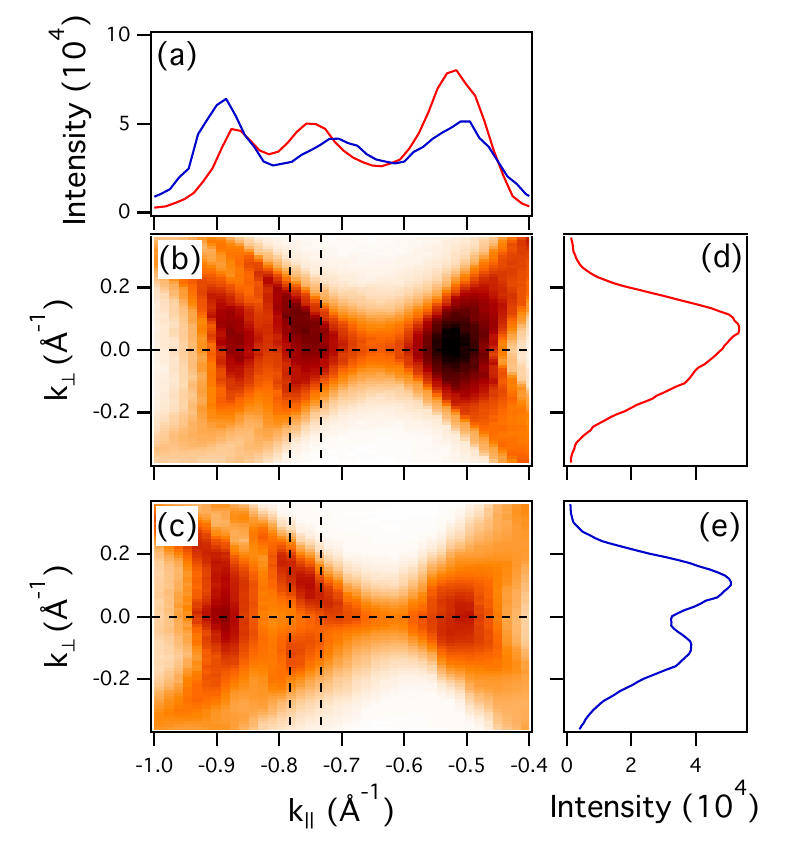}
\caption{Differences between the NS and CDW FSs. (a) Momentum distribution curves (MDCs) along the $\Gamma$-K line of 2H-NbSe$_2$ in the NS ($T=35$ K, red) and CDW ($T=18$ K, blue) state, respectively. (b) FS near the $\Gamma$-K line in the NS. (c) the same as (b), but in the CDW state. (d-e) MDC from the region marked by the two  dashed lines in (b-c).  
}
\label{Fig2}
\end{center}
\end{figure}
%################################################## 

%%##################################################
\begin{figure}[htpb]
\begin{center}
\includegraphics[width=8cm]{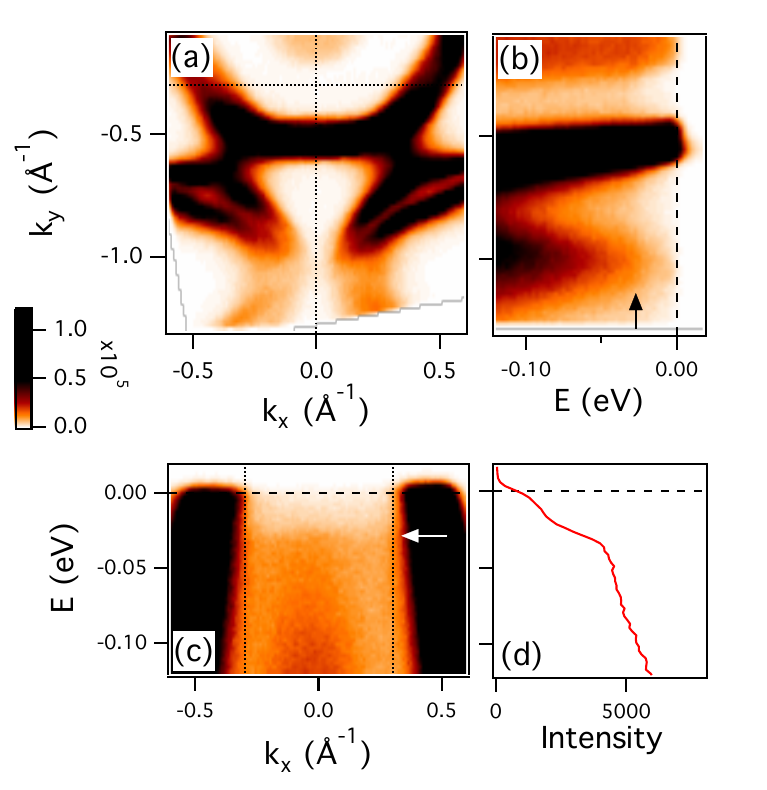}
\caption{Flat electronic feature in 2H-NbSe$_2$. (a) FS and (b, c) dispersion of states along the two dotted lines in (a), on an overexposed false-color scale. (d) An energy distribution curve (EDC) of intensity at $k_y=-0.3$ \AA, integrated over the range of $k_x$ as indicated by dotted lines in (c). The spectra were recorded at 18 K, using the He I radiation. 
}
\label{Fig3}
\end{center}
\end{figure}
%###################################################

%###################################################
\begin{figure*}[htpb]
\begin{center}
\includegraphics[width=16cm]{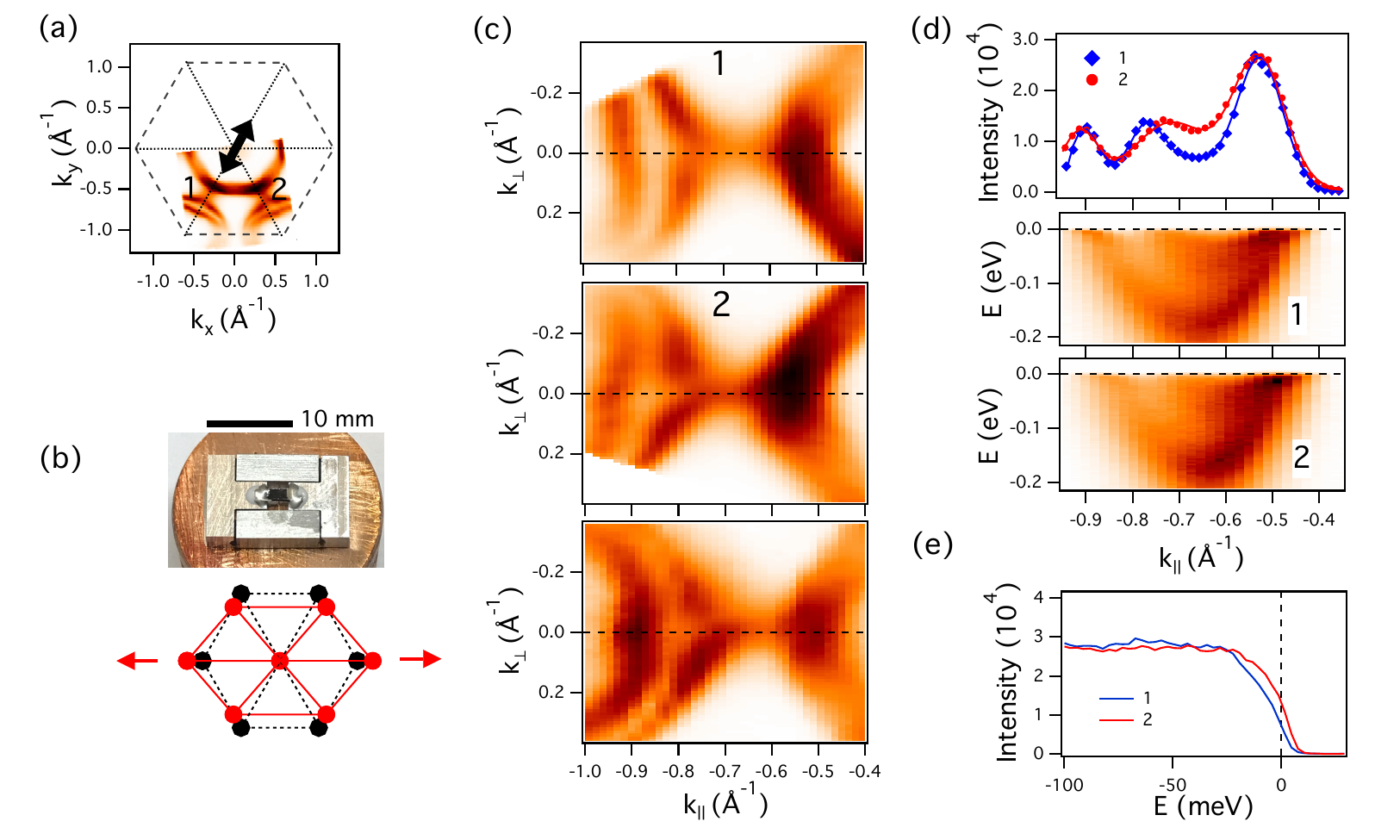}
\caption{Uniaxially strained 2H-NbSe$_2$. (a) The FS of the 2H-NbSe$_2$ sample when the tensile strain of about 0.6\% is applied along the $\Gamma-$K line indicated by an arrow. (b) Sample on the strain device (top) and schematic of the change in the crystal structure of NbSe2 under uniaxial strain (bottom). (c) Close-ups of the regions near the  two $\Gamma-$K lines marked as "1" (top) and "2" (middle) in (a). The same region for the relaxed sample is shown in the bottom panel. (d) Dispersion of states along the two $\Gamma-$K lines marked as "1" and "2" in (a), shown in middle and bottom panel, respectively. The two MDCs at $E=0$ and the corresponding fits are shown in the top panel. (e) The EDCs at the two vHS points at the lines (1) and (2). The strained and relaxed samples were measured at 14 and 18 K, respectively, using He I radiation.
}
\label{Fig4}
\end{center}
\end{figure*}
%################################################## 

\section*{Results}
\subsection*{CDW in relaxed 2H-NbSe$_2$}

Figure \ref{Fig1} illustrates previously unresolved effects of the CDW on the FS of 2H-NbSe$_2$. Panels (a) and (b, d) show the normal state FS recorded with 80 eV and 21.22 eV photons, respectively. The FS of 2H-NbSe$_2$ is formed from the bonding and anti-bonding combinations of the Nb$-4d$ orbitals, resulting in the double-walled barrel-shaped pockets around the K and $\Gamma$ points of the hexagonal Brillouin zone, in a good agreement with the tight binding (TB) approximation \cite{Soumyanarayanan2013}. Earlier ARPES studies have indicated that changes in the electronic structure associated with the CDW transition were very small and often undetectable \cite{Valla2004,Kiss2007,Borisenko2008,Shen2008,Arguello2015}. One possible reason is that in the previous studies, samples would usually be glued to a Cu-sample plate and heated to $\sim100^{\circ}$ C for epoxy to cure. On cooling below the CDW transition, that creates a significant and inhomogeneous bi-axial in-plane compression ($\approx0.4$ \%, based on the thermal expansion coefficients), expected to significantly suppress and smear the CDW. In the present study, this has been avoided by fixing only one corner of the sample, allowing it to fully relax during cooling. As a result, in panels (c), (e) and (f), representing the FS in the CDW state, the discrepancies from the normal state (NS) Fermi barrels centered around K points can be easily identified. First, there are six localized regions of reduced intensity (relative to the normal state FS) as indicated by the dashed circles in Fig. \ref{Fig1}(f). Second, the inner K-centered barrel seems to straighten near the intersections with the zone boundaries, marked by the black arrows. Third, the outer barrel gets additionally bent towards the three pairs of $\Gamma$ centered barrels as indicated by the blue arrows. We note that the FS nesting can not explain these observations. Nonetheless, the observed effects could be interpreted as tendencies for the original K-centered barrels to avoid crossings with the replicas of the FS shifted by the $q_{\mathrm{CDW}}$, as sketched in Fig. \ref{Fig1}(f) (only the replicas shifted by $q_{\mathrm{CDW}}$ along $k_y$ are shown). In reality, however, the replicas are too weak to be observed directly. 
Figure \ref{Fig2} further illustrates spectral changes related to the CDW tranistion.  
The pattern of intensity reduction observed in Figures \ref{Fig1} and \ref{Fig2} would imply that the displacements of Nb atoms in the CDW state are star-shaped, as suggested by calculations, STM and diffraction experiments \cite{Hess1991,Malliakas2013a,Zheng2018}. We note that the intensity reduction is not related to the opening of a real gap, rather, the intensity is partially lost over some energy range, whereas the leading edge is very near the Fermi level. The gap, if it exists, is anomalously small over the whole FS, with the upper limit $<2$ meV.

We also want to point out that there is a weak flat feature, causing a step-like increase in intensity at $\sim30$ meV below the Fermi level, over the large portion of the BZ, as can be seen in Fig. \ref{Fig3}. The indications of a similar feature have been observed in another ARPES study by Bawden \textit{et al} \cite{Bawden2016}.  In the STM studies, an observed partial loss of intensity over the same energy range, $\pm\sim30$ meV, was usually tied to the CDW gap \cite{Wang1990,Hess1990,Soumyanarayanan2013}. However, this feature exists also in the normal state and is definitely not related to the CDW gap. As there are no highly localized states in the band structure calculations, the only viable candidates for that intensity could be the electrons scattered from elsewhere in the BZ due to the EPC, or those (quasi) elastically scattered from the vHSs. The EPC is indeed strong in 2H-NbSe$_2$ and in the presence of defects or impurities, it can contribute to intensity elsewhere in the zone at energies characteristic for the involved phonons \cite{Valla1999a,Valla2006b,Valla2004}. Similarly, the regions of high density of states, such as vHSs will be scattered by defects and will contribute to the intensity at the corresponding energy. Additionally, in the CDW state, the six original vHSs will be replicated, forming an apparent flat band that could become visible, even if the CDW replicas are weak.

\subsection*{Uniaxially strained 2H-NbSe$_2$}

%%##################################################
\begin{figure*}[htpb]
\begin{center}
\includegraphics[width=17cm]{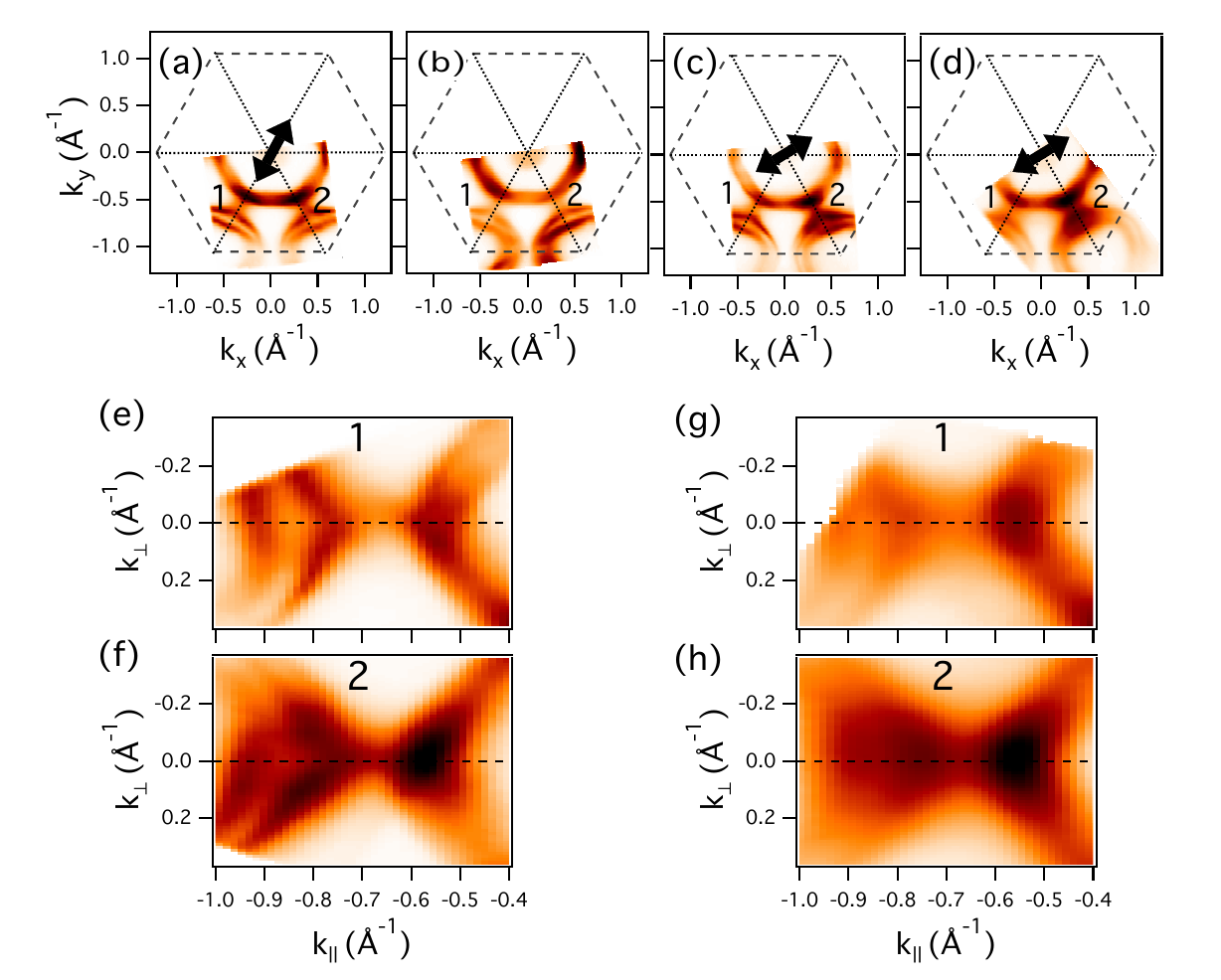}
\caption{Two different directions of uniaxial strain. (a) FS of the 2H-NbSe$_2$ sample under $\approx 0.6\%$ of tensile strain applied along the $\Gamma-$K line, as indicated by the arrow. (b) FS of a relaxed sample. (c) FS of the sample under $\approx 0.6\%$ of tensile strain applied in the $\Gamma-$M line, as indicated by the arrow. (d) FS of the sample in the normal state, under $\approx 0.4\%$ of tensile strain applied in the $\Gamma-M$ line, as indicated by the arrow. The FS maps in (a-d) were recorded at 14, 18, 20 and 100 K, respectively, using the He I radiation. 
(e,f) Close-ups of the regions near the  two $\Gamma-$K lines marked as "1" and "2" from (c). (g,h) The same, but for the normal state (d).}
\label{Fig5}
\end{center}
\end{figure*}
%###################################################

Now, we turn to the effects of uniaxial strain on the electronic sturcture of 2H-NbSe$_2$. Figure \ref{Fig4} shows the changes in the electronic structure of 2H-NbSe$_2$ when a uniaxial tensile strain is applied along the $\Gamma-$ K line in the momentum space (as indicated in panel (a)), corresponding to the stretching the crystal parallel to the Nb-Nb bond (panel (b)). Even though the amount of strain is relatively small, $\sim 0.6\%$, the FS is already visibly distorted: the outside barrels along the tensioned $\Gamma-$ K direction move further apart than those on the initially equivalent $\Gamma-$ K' line. This is further emphasized in the close-ups of the relevant FS segments, shown in Fig. \ref{Fig4}(c). As a consequence, the van Hove singularity of the anti-bonding state, that forms the outer Fermi barrels, gets pulled deeper down along the stretched $\Gamma-$K line, "1", while it gets closer to the Fermi level on the other $\Gamma-$K line, "2", as indicated in Fig. \ref{Fig4}(d,e). Although the energy difference is only $\sim6$ meV in the leading edge, the deformation of the FS is significant, since the vHSs are shallow and the states disperse slowly. The final result is that the global six-fold symmetry from the relaxed state is broken, and the electronic structure acquires 2-fold symmetry. Though the exact amount of strain is not known, our experimental results are consistent with the deformation obtained from calculations for 0.6\% of applied uniaxial strain \cite{Wieteska2019}.

To further test our results and explore the effects of uniaxial strain, we have applied similar amount of tensile strain along the $\Gamma-$M momentum line, as shown in Fig. \ref{Fig5}(c,e,f). We note that in order to keep the volume constant, the stretching along the $\Gamma-$ M line causes compression along the perpendicular $\Gamma-$K line, "2". The resulting compression along the line "2" creates the effects that are very similar to those caused by stretching along the line "1" (Fig.\ref{Fig5}(a) and Fig\ref{Fig4}). The effects are almost indistinguishable, indicating that the Poisson's ratio is nearly $1/2$. Therefore, stretching the crystal along the given $\Gamma-$ K line will shift the vHS on that line deeper and make the distance between the $\Gamma$- and K-centered FSs larger, while the compression along the same line will do the opposite. 

The six-fold symmetry breaking induced by the uniaxial strain is also visible in the normal state, as shown in Fig. \ref{Fig5}(d,g,f). As the temperature at which the map was recorded (100 K) is now closer to room temperature, the strain is expected to be proportionally lower, $\sim 0.4\%$. Again, stretching the sample along the indicated $\Gamma-$ M line causes compression along the  perpendicular $\Gamma-$ K line "2", which in turn brings the corresponding vHS closer to the Fermi level. Therefore, we expect that even the normal state properties should be affected and that the normal state should show an \textit{in-plane} anisotropy under uniaxial strain. We note that there are some differences between straining the crystal along the same line in the CDW state and in the normal state. When comparing Fig. \ref{Fig5}(e,f) with panels (g,h), one can see that the effect on the  distances between the $\Gamma$- and K-centered barrels is very similar. However, the shape of the K-centered barrels is much more symmetric with respect to the $k_{\perp}=0$ line in the normal state (Fig. \ref{Fig5}(g,h)) than in the CDW state (Fig. \ref{Fig5}(e,f) and Fig. \ref{Fig4}), with no obvious skewing of the inner K-centered barrel.

\section*{Discussion}
We note that the applied distortion ($0.4-0.6 \% $) is too small to observe a deformation of the Brillouin zone in ARPES. However, such a small distortion of the lattice caused detectable changes in the electronic structure, as shown in Fig. \ref{Fig4} and Fig. \ref{Fig5}. This is due to the significant and highly anisotropic change in the hopping parameters that a small lattice distortion causes. Of particular importance is the demonstrated ability to move the vHSs on the $\Gamma-$K lines by strain. If brought exactly to the Fermi level, one could expect an enhancement of correlations due to increased DOS, and an increase in the superconducting transition temperature. 

\begin{figure}[htpb]
\begin{center}
\includegraphics[width=8.5cm]{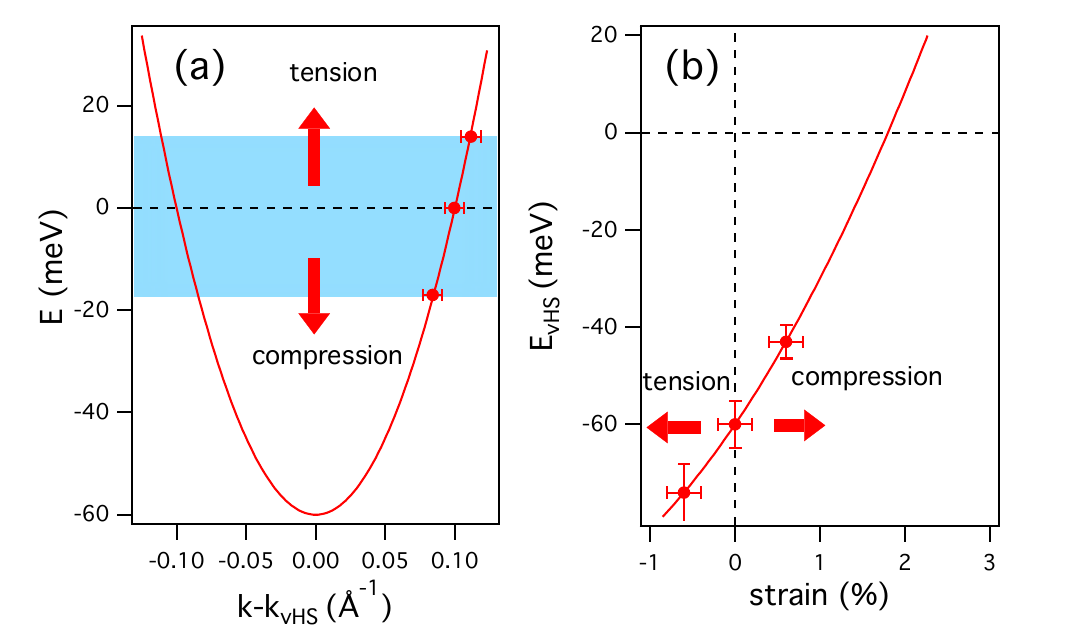}
\caption{The energy of vHS vs. strain. (a) Dispersion of antibonding state along the $\Gamma-$ K line obtained by approximating it with parabola with the minimum taken from the TB approximation, $E_{\mathrm{vHS}}=-60$ meV, and the measured values for $k_{\mathrm{F}}$, for the relaxed state and two strained cases from Fig. \ref{Fig4}. The error bars correspond to the standard deviation (SD) of the fitted MDC peak positions from Fig. \ref{Fig4}(d). (b) Distance of the vHS from the Fermi level as modeled in (a). The $x$ and $y$ error bars correspond to the estimated uncertainty in the strain ($\pm{0.2}\%$) and the SD in energy due to uncertainty in $k_{\mathrm{F}}$ from (a), respectively.
}
\label{Fig6}
\end{center}
\end{figure}

Even though our strain device does not allow the strain to be varied at constant temperature, we can give a rough estimate of the strain dependence of the energy of vHS. The dispersion of the antibonding state along the $\Gamma-$ K line (Fig. \ref{Fig4}(d)) can be crudely approximated by a parabola going through the bottom at $E_{\mathrm{vHS}}=-60$ meV and two $k_F$ points. Since the state is much sharper at the Fermi level than at the bottom, the changes due to strain are easier to measure through $k_{\mathrm{F}}$. This is illustrated in Fig. \ref{Fig6}. If we keep the mass constant and take into account the measured changes in $k_{\mathrm{F}}$ under strain, we can estimate the strain required to bring the vHS to the Fermi level to roughly 1.8 \%. We note that the slowly dispersing leading edge of the state in Fig. \ref{Fig4}(d,e) shifts much less than vHS in Fig. \ref{Fig6}. This is due to the strong renormalization of the state's dispersion caused by the EPC. Its shift under strain represents a strong indication that the EPC strengthens and that the involved phonons soften as the vHS gets closer to the Fermi level when the given $\Gamma-$ K line is compressed.

%%##################################################
%###################################################
There are other, more subtle effects, related to the signatures of CDW that we discussed previously for the relaxed sample. It appears that the strength of CDW signatures is slightly different on the two $\Gamma-$K lines and on the two sets of K-centered barrels when the strain is applied. For example, it seems that the bending of the outer Fermi barrels, marked by the blue arrows in Fig. \ref{Fig1}(f), is stronger when the barrels are closer (Fig. \ref{Fig4}). Additionally, in the CDW state, the inner K-centered barrels get skewed, particularly those that are not on the lines of new symmetry introduced by the strain direction. As already noted, this is not observed in the normal state, implying that the CDW state is more susceptible to uniaxial strain. This could be related to additional symmetry breaking where the CDW tends to transition from a 3Q to 1Q state under uniaxial strain \cite{Gao2018,Soumyanarayanan2013,Wieteska2019,Flicker2015}.  
In the relaxed state, the $\Gamma-$ K lines and the K-center barrels are all equivalent, both in the normal state and in the CDW state (3Q). Therefore, our results indicating inequivalency in the strained state would imply that the CDW state itself should be strongly affected. This has indeed been observed in recent STM studies where, under an inhomogeneous bi-axial strain, local stripy regions (1Q) compete with the 3Q order characteristic of the relaxed state \cite{Gao2018,Soumyanarayanan2013}. 

Depending on the applied direction, the uniaxial strain could enhance one of the 3 CDWs, while weakening the other two. This should also break the symmetry of phonon branches and should affect the EPC and superconductivity, too. Just the fact that the positions of vHSs are affected by the uniaxial strain would indicate that SC should be affected, analogously to the Sr$_2$RuO$_4$ case. Here, the situation is expected to be even more interesting due to the co-existence of the two orders. It would be interesting  to check if, and how the transition temperatures of SC and CDW are affected. Therefore the comprehensive macroscopic properties studies (specific heat, magneto-transport, magnetization etc.) under controlled uniaxial strain would be highly desirable. 

In summary, our studies have shown that the initial six-fold symmetry of the electronic structure is visibly broken if a uniaxial strain of ~0.6\% is applied on a bulk crystal. Depending on the applied direction, the uniaxial strain would likely enhance one of the 3 CDWs, while weakening the other two, or vice versa. 
This should also have consequences on correlations, including the EPC, self-energy, gaps and lattice modes and should affect both the CDW and SC phases and alter the macroscopic properties of the material. We hope that our study will stimulate highly needed studies of those properties under uniaxial strain. 

\section*{Methods}

The experiments within this study were done in the experimental facility that integrates oxide molecular beam epitaxy synthesis with ARPES and STM capabilities within the common vacuum system \cite{Kim2022}. The samples intended for studies of the relaxed state were glued to the Cu sample plate by Ag-epoxy only at one corner. The samples for uniaxial strain studies were glued over the strain bridge, without substrate, by a combination of stycast 2850 and Ag-epoxy in such a way that two corners of the  sample were glued, while the strained part was suspended over the gap. The sample was then cleaved with Kapton tape in the ARPES preparation chamber (base pressure of $3\times10^{-8}$ Pa). The cleaved samples were then transferred to the ARPES chamber (base pressure of $8\times10^{-9}$ Pa). 
The strain device uses the differential thermal contraction between aluminium and titanium to uniaxially tension the sample. This differential contraction produces up to 0.8 \% of tensile strain on cooling from room temperature to $\sim10$ K \cite{Sunko2019a}. 

The ARPES experiments were carried out on a Scienta SES-R4000 electron spectrometer with the monochromatized HeI (21.22 eV) radiation (VUV-5k). The total instrumental energy resolution was $\sim$ 3 meV, as measured by the width of the Fermi level from the cold spot on the cryostat ($T\leq 6$ K). Temperature of the sample was then determined by the width of the Fermi level, accounting for the instrumental energy resolution. Synchrotron studies were performed at the ESM beamline by using $h\nu=80$ eV and total experimental energy resolution was $\sim$ 12 meV. Angular resolution at both facilities was better than $\sim 0.15^{\circ}$ and $0.3^{\circ}$ along and perpendicular to the slit of the analyzer, respectively.

\section*{Data availability} 
The data that support the findings of this study are available from the corresponding author upon reasonable request. %The source data underlying 6a-b are provided as a Source Data file.

\section*{Acknowledgements}

This research used ESM beamline of the National Synchrotron Light Source II, a U.S. Department of Energy (DOE) Office of Science User Facility operated for the DOE Office of Science by Brookhaven National Laboratory under Contract No. DE-SC0012704.
T.V. and I.I.K. acknowledge the support from the Red guipuzcoana de Ciencia, Tecnología e Innovación – Gipuzkoa NEXT 2023 from the Gipuzkoa Provincial Council under Contract No. 2023-CIEN-000046-01.

\section*{Author information}
\subsection*{Affiliations}

Condensed Matter Physics and Materials Science Department, Brookhaven National Laboratory, Upton, New York 11973, USA.\\
T. Valla, A. K. Kundu\\
National Synchrotron Light Source II, Brookhaven National Laboratory, Upton, New York 11973, USA.\\
A. K. Kundu, A. Rajapitamahuni, E. Vescovo\\
Donostia International Physics Center, 20018 Donostia-San Sebastian, Spain.\\
I. I. Klimovskikh, T. Valla\\
Institute of Condensed Matter Physics, Ecole Polytechnique Fédérale de Lausanne, 1015 Lausanne, Switzerland.\\
H. Berger\\

\subsection*{Author Contributions}
T.V. designed and directed the study and wrote the manuscript. A.K.K, I.I.K. and T.V. performed the ARPES experiments and analyzed and interpreted data. A.R. and E.V. assisted with the synchrotron ARPES experiments. H.B. grew the single-crystals. 

\subsection*{Competing interests}
The authors declare no competing interests.\\

\subsection*{Corresponding author}
Correspondence should be addressed to T.V. (tonica.valla@dipc.org).

%

%\bibliography{NbSe2}

\end{document}